\documentstyle[12pt,amstex,amssymb,righttag]{article}

\begin{document}
\renewcommand{\thefootnote}{\fnsymbol{footnote}}

\thispagestyle{empty}

\vspace*{-1cm}
\begin{center}
{\Large \bf Non Parallels Electric and Magnetic Fields in a \\
\vspace*{-3mm}
F.R.W. Cosmology. Classical and Quantum\\
Gravitational Implications}

\vspace{3mm}
by\\
\vspace{3mm}
{\sl Carlos Pinheiro$^1$\footnote{fcpnunes@@cce.ufes.br/maria@@gbl.com.br}, J.A. Hela\"{y}el-Neto$^{\ast \ast}$
and Gilmar S. Dias$^{\ast \ast \ast}$
} 

\vspace{3mm}
$^{\ast}$Universidade Federal do Esp\'{\i}rito Santo, UFES.\\
Centro de Ci\^encias Exatas\\
Av. Fernando Ferrari s/n$^{\underline{0}}$\\
Campus da Goiabeiras 29060-900 Vit\'oria ES -- Brazil.\\

\noindent
$^{\ast \ast}$Centro Brasileiro de Pesquisas -- CBPF/CNPq\\
Rua Dr. Xavier Sigaud, 150\\
22290-180 - Rio de Janeiro, RJ -- Brasil\\
Universidade Cat\'olica de Petr\'opolis -- UCP\\

\noindent
$^{\ast \ast \ast}$Departamento de F\'{\i}sica -- UFES\\
Escola T\'ecnica Federal, ETFES
\end{center}

\vspace{3mm}
\begin{center}
Abstract
\end{center}
At first, we discuss parallels electric and magnetic fields solutions
in a gravitational background. Then, considering eletromagnetic and
gravitational waves symmetries we show a particular solution for
stationary gravitational waves. Finally we consider gravitation as a
gauge theory (effective gravitational theory), evaluate the
propagators of the 
model, analyze the corresponding quantum excitations and verify
(confirm) the tree-level unitarity -- at many places of the model.

\newpage
\section{Introduction}
\setcounter{footnote}{0}
\paragraph*{}

The physical and mathematical aspects of the vector equations 
$\vec{\nabla}\times \vec{V}=k\vec{V}$($\vec{V}$ is a vector field and $k$
is a positive constant) have frequently been analyzed. Particularly,
plasma and astrophysical plasma physics has pointed out the existence
of stationary eletromagnetic waves that seems to be the most important
aspect of this equation. Particular solutions have been found for the
classical electrodynamics vector potential with the form [1-8]
\begin{equation}
\vec{A}=a[i\ \sin kz+j\cos kz]cos \omega t
\end{equation}

The associated electric and magnetic fields are:
\begin{eqnarray}
\vec{E} &=& -\frac{1}{c}\ \frac{\partial \vec{A}}{\partial t}=ka[i\sin kz+
\hat{j}\cos kz]
\sin\omega t\\
\vec{B} &=& \vec{\nabla}\times \vec{A}=ka[i\sin kz +j\cos kz]\cos \omega t
\end{eqnarray}
The fields (2) and (3) satisfy Maxwell equations, as well as, the
usual vacuum free-wave equation. Moreover, fields $\vec{E}$, $\vec{B}$
and $\vec{A}$ are parallel everywhere. It is interesting to point
out, these electromagnetic waves do not propagate energy, having a
null Poynting vector. From the $\vec{\nabla} \times \vec{A}=k\vec{A}$, we
get the equation:
\begin{equation}
\nabla^2\vec{A}+k^2\vec{A}=0
\end{equation}
with equation (1) as a possible solution. 

We argue that if astrophysical plasma is considered, there must be
some restrictions to get the parallel $\vec{A}$, $\vec{E}$ and
$\vec{B}$ fields solutions reported by Brownstein et. al. \cite{tres,oito}.

This work shows that it is not always possible to get these solutions
if a strong gravitational background is considered.
Gravitation breaks the fields parallelism. So the associated
eletromagnetic wave does not have a null Poynting vector anymore and
propagates energy in a gravitational background. Then, looking for
equation like (1.4) for gravitational field and a particular solution,
we analyze of a possible stationary gravitational wave. It
could be new gravitational waves, different from the general
relativity gravitational waves (taken from linearization of Einstein
equations). Experimentally, this stationary gravitational waves may be
found in black-hole distributions.

Finally, using electromagnetic and gravitational gauge theory
symmetries we analyse stationary quantum gravitational wave. Here we
consider gravitation as a $4$-dimensional effective theory. 

\section{Parallel $\vec{A}$, $\vec{E}$ and $\vec{B}$ fields in Plasma
and Astrophysical Plasma}

\paragraph*{}
The fields $\vec{A}$, $\vec{E}$ and $\vec{B}$ are parallel in vaccum
or, under some conditions, in plamas, what should follow if a
gravitational background is taken into account? Do Brownstein fields
remain parallel? and does the associated electromagnetic wave
propagate energy?

In order to check these questions we consider a gravitational and
electromagnetic coupling by the action:
\begin{equation}
{\cal S}=\int\sqrt{-g}\left(-\ \frac{1}{4}\ F_{\mu \nu}F^{\mu \nu}\right)d^4x
\end{equation}

In this gravitational background the Maxwell inhomogeneous equations are:
\begin{equation}
{\cal D}_{\mu}F^{\mu \nu}={\cal J}^{\nu}
\end{equation}
where 
\begin{equation}
{\cal D}_{\mu}F^{\mu \nu}=\partial_{\mu}F^{\mu \nu}+\Gamma^{\mu}_{\mu
\lambda}F^{\lambda \nu}+\Gamma^{\nu}_{\mu \lambda}F^{\mu \lambda}=
{\cal J}^{\nu}
\end{equation}
and Maxwell homogeneous equations are:
\begin{equation}
{\cal D}_{\mu}F_{\nu \rho}+{\cal D}_{\nu}F_{\rho \mu}+{\cal
D}_{\rho}F_{\mu \nu}=0
\end{equation}

The connections terms cancel each other such that this last equation is the
usual Maxwell homogeneous equation:
\begin{equation}
\partial_{\mu}F_{\nu \rho}+\partial_{\nu}F_{\rho \mu}
+ \partial_{\rho}F_{\mu \nu}=0\ .
\end{equation}

To solve inhomogeneous equations (8) we adopt the F.R.W. cosmological
metric:
\begin{equation}
dS^2=dt^2-a^2_{(t)}\left\{\left(1-Ar^2\right)^{-1}dr^2+r^2d\theta^2+
r^2\sin^2\theta d\varphi^2\right\}
\end{equation}

At the same time, the second term of the covariant derivative
equation (7) may be written appropriately, with $\Gamma^{\mu}_{\mu
\lambda}$ given as 
\begin{equation}
\Gamma^{\mu}_{\mu \lambda}=-\ \frac{\partial}{\partial x^{\lambda}}\ \log
\sqrt{-\tilde{g}}\ ,
\end{equation}
where $\tilde{g}$ is the metric determinant.

Then Maxwell equations (7) and (8) are explicitly given by:
\begin{equation}
\vec{\nabla}\cdot \vec{E}-g\vec{\nabla}f\cdot
\vec{E}=\rho_{(\vec{x})}\ ,
\end{equation}
\[
\vec{\nabla}\cdot \vec{B}=0\ ,
\]
\[
\vec{\nabla}\times \vec{E} =-\ \frac{\partial \vec{B}}{\partial t}\ ,
\]
\[
\vec{\nabla}\times \vec{B}=\vec{J}(\vec{x})+\frac{\partial \vec{E}}
{\partial t}-g\ \frac{\partial f}{\partial t}\ \vec{E}+g\vec{\nabla}f\times \vec{B}-
\Gamma^i_{\mu \lambda}F^{\mu \lambda}\ .
\]

Without electromagnetic sources, that is, for $\rho =0$ and
$\vec{J}=0$ we get ``free'' electromagnetic fields equations in a
graviational background:
\begin{equation}
\vec{\nabla}\cdot \vec{E}=g\vec{\nabla}f\cdot \vec{E}\ ,
\end{equation}
\begin{equation}
\vec{\nabla}\cdot \vec{B}=0\ ,
\end{equation}
\begin{equation}
\vec{\nabla}\times \vec{E}=-\ \frac{\partial \vec{B}}{\partial t}\ ,
\end{equation}
\begin{equation}
\vec{\nabla}\times \vec{B}=\frac{\partial \vec{E}}{\partial t}-
g\ \frac{\partial f}{\partial t}\vec{E}+g\vec{\nabla}f\times
\vec{B}-\Gamma^i_{\mu \lambda}F^{\mu \lambda}
\end{equation}
where. $g=\displaystyle{\frac{\sqrt{1-Ar^2}}{a^3r^2\sin \theta}}$ and
$f=\displaystyle{\frac{a^2r^2\sin \theta}{\sqrt{1-Ar^2}}}$.

Functions $g$ and $f$ may have $A=+1,0,-1$ each value represents the
associated curvature of F.R.W. spatial metric section.

The electric field wave equation in a gravitational background is
obtained from eq. (15) as 
\begin{eqnarray}
\nabla^2\vec{E} &-& \frac{\partial^2\vec{E}}{\partial t^2}=
\vec{\nabla}\left(g\vec{\nabla}f\cdot \vec{E}\right)-
\frac{\partial g}{\partial t}\ \frac{\partial f}{\partial t}\ 
\vec{E}-g \ \frac{\partial ^2f}{\partial t^2}\ \vec{E}-
g\ \frac{\partial f}{\partial t}\ \frac{\partial \vec{E}}{\partial t}+
\\
&+& \frac{\partial g}{\partial t}\ \vec{\nabla}f\times \vec{B}+
g\ \frac{\partial}{\partial t}\ \vec{\nabla}f\times 
\vec{B}+g\vec{\nabla}f\times \frac{\partial \vec{B}}{\partial t}-\frac{\partial}{\partial t}\ 
\left(\Gamma^i_{\mu \beta}F^{\mu \beta}\right)\ .\nonumber
\end{eqnarray}

The magnetic field wave equation in a gravitational background is
obtained from eq. (16) as 
\begin{equation}
\nabla^2\vec{B}-\frac{\partial^2\vec{B}}{\partial t^2}=
\vec{\nabla}\times \left(g\ \frac{\partial f}{\partial t}\ \vec{E}-
g\vec{\nabla}f\times \vec{B}+\Gamma^i_{\mu \beta}F^{\mu
\beta}\right)\ .
\end{equation}

The last term, $\Gamma^{\nu}_{\mu \lambda}F^{\mu \lambda}$, is 
identically zero. But we still retain it for esthetic completeness of
equation (2.7). At the same time, equations (2.17) and (2.18) reproduce the correponding vacuum equations
if gravitation is ignored.

Using equation (2.17) and (2.18) it can be shown that there is at least, one
solution in which fields $\vec{E}$ and $\vec{B}$ are not parallel
anymore. For example, putting field $\vec{B}$ as given by (2.3), in
equation (2.18), and writting the gradient $\vec{\nabla}f$ as:
\begin{equation}
\vec{\nabla}f=\hat{r}_0\frac{\partial f}{\partial r}+\hat{\theta}_0
\frac{\partial f}{\partial \theta}+\hat{\varphi}_0\ 
\frac{\partial f}{\partial \varphi}\ ,
\end{equation}
it is easy to verify that equation (2.18) becomes
\begin{equation}
\frac{\partial f}{\partial t}\ \vec{E}=\vec{\nabla}f\times \vec{B}
\Longrightarrow \vec{E}=\frac{\vec{\nabla}f\times \vec{B}}
{\displaystyle{\frac{\partial f}{\partial t}}}
\end{equation}

Calculating $\displaystyle{\frac{\partial f}{\partial t}}$, 
$\vec{\nabla}f$, $\vec{\nabla}f\times \vec{B}$ and taking explicitly
the unity vectors $\hat{r}_0$ and $\hat{\theta}_0$
\begin{eqnarray}
\hat{r}_0 &=& \hat{i}\sin \theta \cos \varphi +\hat{j}\sin \theta \sin
\varphi + \hat{k}\cos \theta \\
\hat{\theta}_0 &=& \hat{i}\cos \theta \cos \varphi 
+\hat{j}\cos \theta \sin \varphi -\hat{k}\sin \theta  \nonumber 
\end{eqnarray}
we can find, 
\begin{equation}
\hat{B}=ka\left[i\sin  (kz)+\hat{j}(kz)\right]\cos (\omega t)\ ,
\end{equation}
\begin{eqnarray}
\vec{E} &=& \hat{i}\left(\sin \theta G_{(r,t,\theta )}-\cos \theta F_{(r,t)}\right)
ka\cos (kz)\cos (\omega t)+\\
&+& \hat{j}\left(\cos \theta F_{(r,t)}-\sin \theta G_{(r,t,\theta )}\right)ka
\sin (kz)\cos (\omega t)+\nonumber \\
&+& \hat{k}\left[\left(\sin \theta \cos \varphi F_{(r,t)}+\cos \theta
\cos \varphi G_{(r,t,\theta )}\right)ka\cos (kz)\cos (\omega t)+\right.\nonumber \\
&-& \left.\left(\sin \theta \sin \varphi F_{(r,t)}\cos \theta \sin
\varphi G_{(r,t,\phi )}\right)ka\sin (kz)\cos (\omega t)\right]\nonumber  
\end{eqnarray}
where functions $F_{(r,t)}$ and $G_{r,t,\theta )}$ are given by:
\begin{eqnarray}
F_{(r,t)} &=& \frac{2a}{3\dot{a}r}+\frac{Aar}{3\dot{a}(1-Ar^2)}\quad \mbox{and}\nonumber \\
G_{(r,t,\theta )} &=& \frac{a\mbox{cotg} \theta}{3\dot{a}r}\ .\nonumber 
\end{eqnarray}

The new electrital field $\vec{E}$ given by (2.23) is completely
different from the Brownstein electrical field (1.2). Now, fields
$\vec{E}$ and $\vec{B}$ are perpendicular vectors, as we can see from
equation (2.20). Gravitational backgound breaks $\vec{E}$ and $\vec{B}$
parallelism and so the Poynting vector $\vec{S}=\vec{E}\times
\vec{B}$ is not zero anymore.

This way, the associated electromagnetic wave propagates energy.

It is possible the electric field as the Brownstein electric field 
given in eq. (1.2) is kept and a new particular solution for magnetic $\vec{B}$
field in a gravitational background is obtained. We believe, this new
magnetic field differs from the Brownstein magnetic field
given in eq. (1.3) and the
associated electromagnetic wave propagates energy.

It is not always possible to have $\vec{E}$, $\vec{B}$ and $\vec{A}$
as parallel fields if a gravitational background is taken into
account. On the other hand, even in a gravitational background, it
might be possible to claim parallel $\vec{E}$ and $\vec{B}$ fields,
writing $\vec{E}=\chi \vec{B}$ in equations (2.17) and (2.18) and to get
conditions that the function $\chi$ must satisfy. 

\section{Stationary Gravitational Waves}\setcounter{equation}{0}

\paragraph*{}
Now, considering the similarities of electromagnetic, gravitational
and linearized Einstein gravitation theories, we analyze a
possibility of stationary gravitational wave.

According to the phenomenological view, black-hole distributions may
be nodes of gravitational waves propagating in space. It would be
sufficient for two black-holes separated by large distances,
if the space-time curvature of the strong gravitational fields of
the black-holes is neglected. So the space-time of one black-hole
is assymptotically flat to the other. Then each black-holes is
like a `string node' and this mechanism may confine gravitational waves.

The following equation may reproduce these waves in nature:
\begin{equation}
R_{\mu \nu}=\kappa \Lambda h_{\mu \nu}
\end{equation}
where 
\begin{equation}
g_{\mu \nu}=\eta_{\mu \nu}+\kappa h_{\mu \nu}\ .
\end{equation}

Here $\kappa$ is the Newton gravitational constant, $\Lambda$ is
the cosmological constant, $h_{\mu \nu}$ is the stationary
gravitational perturbation in a Minkowski ``Background''. Then the
stationary gravitational pertubation is given by 
\begin{equation}
\partial_{\beta}\partial_{\nu}h^{\beta}_{\mu}+
\partial_{\beta}\partial_{\mu}h^{\beta}_{\nu}-
\Box h_{\mu \nu}-\partial_{\mu}\partial_{\nu}h^{\beta}_{\beta}=
\Lambda h_{\mu \nu}\ .
\end{equation}

This equation is formally similar to the electromagnetic case,
eq. (1.4).

This equation may have a solution like
\begin{equation}
h_{\mu \nu}=C_{\mu \nu}(z)f(t)\ ,
\end{equation}
where
\begin{equation}
h_{\mu \nu}=
\left[
\begin{array}{cccc}
A_{00} & 0 & 0 & 0\\
0& A_{11} & A_{12} & 0\\
0 & A_{12} & -A_{11} & 0\\
0 & 0 & 0 & A_{00}\\
\end{array}
\right]e^{i\tilde{k}z}\cos \omega t\ ,
\end{equation}
where $A_{00}$, $A_{11}$ and $A_{12}$ are free constants and 
$\tilde{k}=\sqrt{\Lambda -\omega^2}$ is the wave-vector for the case
$\Lambda -\omega^2>0$.

The dispersion relation suggests a long wave-length, $\lambda$, for
stationary gravitational waves in nature, since its frequency is small.

The lagrangean for the pertuberd field is 
\begin{eqnarray}
{\cal L}^{(\Lambda )}_h &=& -\ \frac{1}{4}\ \partial_{\rho}h_{\mu \nu}
\partial^{\rho}h^{\mu \nu}+\frac{1}{4}\ 
\partial_{\rho}h\partial^{\rho}h-\ 
\frac{1}{2}\ \partial_{\rho}h^{\rho \mu}\partial_{\mu}h +\
\frac{1}{2}\ \partial^{\rho}h_{\rho \mu}\partial_{\nu}h^{\mu
\nu}+\nonumber \\
&-& \frac{1}{4}\ \Lambda h^{\mu \nu}h_{\mu \nu}+
\frac{1}{8}\ \Lambda h^2\ .
\end{eqnarray}

>From this lagrangean we get the energy-momentum tensor $T_{\mu \nu}$
and, using the equation of motion with 
$\partial_{\mu}T^{\mu \nu}=0$. The stress tensor is not symmetric.
Then $T_{\mu \nu}$ is 
\begin{eqnarray}
T_{\mu \nu} &=& \frac{1}{4}\ \partial_{\mu}\partial_{\nu}h-\ 
\frac{1}{2}\ \partial_{\mu}h_{\alpha \beta}\partial_{\nu}h^{\alpha
\beta}-\ 
\frac{1}{2}\ \partial_{\nu}h_{\mu \beta}\partial^{\beta}h-\
\frac{1}{2}\ \partial^{\beta}h_{\mu \beta}\partial_{\nu}h+\nonumber \\
&+& \partial^{\beta}h_{\beta \sigma}\partial_{\nu}h^{\sigma}_{\mu}-
\eta_{\mu \nu}{\cal L}^{(\Lambda)}_h\ .
\end{eqnarray}

A system of black-hole distribution may confine stationary
gravitational waves.

\section{Gravitation as a Gauge Theory-Gravitons}

\paragraph*{}
Now, we intend to attack the last problem according to quantum view.
Gravitation is considered a gauge theory.

Classically, stationary gravitational waves between two points $x$ and $y$
have been obtained. According to the quantum view, this problem may
have graviton exchanges. The situation is similar to that of point charges
interacting by photon exchanges.

This way, the quantum version of the problem of stationary
gravitational waves between two points $x$ and $y$ is now described by
graviton creation and annihilation at these same points.

For convecinece, we parametrize the field $h_{\mu \nu}$ as
\[
H^{\alpha}_{\nu}=h^{\alpha}_{\nu}-\ \frac{1}{2}\ 
\delta^{\alpha}_{\nu}h\ .
\]

It is simple to verify that $H^{\alpha}_{\alpha}=H=-h$ where  $h$ is 
the trace of $h_{\mu \nu}$. With this new $h_{\mu \nu}$, the
Lagrangian (29) may be written as
\begin{eqnarray}
{\cal L}^{(\Lambda )}_H &=& \frac{1}{2}\ \partial_{\rho}H_{\mu \nu}
\partial^{\rho}H^{\mu \nu}+\frac{1}{4}\ \partial_{\rho}H\partial^{\rho}H-\
\frac{1}{2}\ \partial^{\rho}H_{\rho \nu}\partial_{\mu}H^{\mu \nu}+
\frac{1}{2}\ \partial^{\rho}H_{\rho \mu}\partial_{\nu}H^{\mu
\nu}+\nonumber \\
&-& \frac{1}{2}\ \Lambda H^{\mu \nu}H_{\mu \nu}+\frac{1}{4}\
\Lambda H^2\ .
\end{eqnarray}

Finally, we integrate the lagrangean (31) by parts and get:
\begin{eqnarray}
{\cal L}^{(\Lambda )}_H &=& \frac{1}{2}\ H^{\mu \nu}\Box H^{\mu
\nu}-\ 
\frac{1}{4}\ H\Box H-\ \frac{1}{2}\ H^{\mu \nu}
\partial_{\mu}\partial_{\alpha}H^{\alpha}_{\nu}-\
\frac{1}{2}\ H^{\mu
\nu}\partial_{\nu}\partial_{\alpha}H^{\alpha}_{\nu}+ 
\nonumber \\
&-&\frac{1}{2}\ \Lambda H^{\mu \nu}H_{\mu \nu}+\frac{1}{4}\ \Lambda
H^2\ .
\end{eqnarray}

Then, it is possible to write this expression in a bilinear form:
\begin{equation}
{\cal L}^{(\Lambda )}_H=\frac{1}{2}\ H^{\mu \nu}\Theta_{\mu \nu
,\kappa ,\lambda}H^{\kappa ,\lambda}\ ,
\end{equation}
where operator $\Theta_{\mu \nu ,\kappa \lambda}$ has the following
form in terms of Barnes-Rivers \cite{nove} spin projection operators:
\begin{eqnarray}
\Theta_{\mu \nu ,\kappa \lambda} &=& (\Box -\Lambda )P^{(2)}-
\Lambda P^{(1)}_m+\frac{5}{2}\ (\Box -\Lambda )P_s^{(0)}-\ 
\frac{(\Lambda +3\Box )}{2}\ P^{(0)}_{\omega}\\
&+& \frac{\sqrt{3}}{2}\ (\Lambda -\Box )P^{(0)}_{\omega}+
\frac{\sqrt{3}}{2}\ (\Lambda -\Box )P^{(0)}_{\omega s}\ .\nonumber
\end{eqnarray}

The inverse operator $\Theta^{-1}_{\mu \nu ,\kappa \lambda}$
has the following form:
\begin{equation}
\Theta^{-1}_{\mu \nu ,\kappa \lambda}=\left[XP^{(2)}+
YP^{(1)}_m+ZP^{(0)}_s+WP^{(0)}_{\omega}+RP_{sw}^{(0)}+
SP^{(0)}_{\omega s}\right]_{\mu \nu ,\kappa \lambda} \ .
\end{equation}

To get the coefficients $X,\ Y\ \cdots S$ it is sufficient to use
Barnes-Rivers \cite{nove} projection operators and its multiplicative
table, such that:
\begin{equation}
\Theta^{\rho \sigma}\mbox{}_{\mu \nu}\Theta^{-1}_{\rho \sigma ,\kappa \lambda}=
(I)_{\mu \nu ,\kappa
\lambda}=\left(P^{(2)}+P^{(1)}_m+P^{(0)}_s+P^{(0)}_{\omega}
\right)_{\mu \nu ,\kappa \lambda}\ .
\end{equation}

For $D=4$ these coefficients have the following form 
\begin{equation}
X=-\ \frac{1}{\Lambda -\Box}\ , \quad
Y=-\ \frac{1}{\Lambda}\ , \quad 
Z=-\ \frac{\Lambda +3\Box}{\Lambda^2+8\Lambda \Box -9\Box^2}\ ,
\end{equation}
\[
W=-\ \frac{5}{\Lambda -9\Box}\ , \quad
R=-\ \frac{\sqrt{3}}{\Lambda +\Box}\ , \quad 
S=-\ \frac{\sqrt{3}}{\Lambda +9\Box}\ .
\]

We get the corresponding propagator by the following functional generator:
\[
W[T_{\rho \sigma}]=-\ \frac{1}{2}\int d^4xd^4yT^{\mu \nu}
\Theta^{-1}_{\mu \nu ,\kappa \lambda}T^{\kappa \lambda}\ .
\]

The propagator is written explicitly as:
\begin{equation}
\langle T\left[h_{\mu \nu (x)};h_{\kappa \lambda (x)}\right]\rangle =
i\Theta^{-1}_{\mu \nu ,\kappa \lambda}\delta^4(x-y)\ .
\end{equation}

\section{Unitarity of the Model}

\paragraph*{}
Now we discuss the tree-level unitarity of the model. Coupling the
propagator and external current, $T^{\mu \nu}$ we analyze the poles of
the current amplitude and the imaginary part of its residue. The
current amplitude is given by:
\begin{equation}
{\cal A}=T^{\ast \mu \nu}(\vec{k})\langle T
\left[h_{\mu \nu}(-\vec{k});h_{\kappa \lambda}(\vec{k})\right]\rangle
T^{\kappa \lambda}(-\vec{k})\ .
\end{equation}

Using equations (4.12), (4.14) and (4.15) we get the current amplitude. And
considering the imaginary part of the residue at the poles, it is simple to
verify that, due to the transverse condition, only operators $P^{(2)}$
and $P^{(0)}_s$ remain. And that it is a symmetry, not necessarily a
gauge symmetry:
\begin{equation}
\omega_{\mu \nu}T^{\mu \nu}=0\ .
\end{equation}

Analyzing the poles of the sector with spin-2 and the sector with spin-$0$ in
the momentum space we get:
\begin{equation}
X=-\ \frac{1}{k^2+\Lambda}
\end{equation}
\[
Z=\frac{3k^2-\Lambda}{\Lambda^2-8\Lambda k^2-9(k^2)^2}\ .
\]

This way we have for the sector with spin-$2$, a tachyon pole:
\begin{equation}
k^2=-\Lambda \ 
\end{equation}
and for the sector with spin-$0$, we have two poles, one being a
tachyon pole
\begin{eqnarray}
k^2 &=& \frac{1}{9}\ \Lambda \ , \\
k^2 &=& -\Lambda \ . \nonumber 
\end{eqnarray}
Using (5.16) and considering the $k^2$ pole analysis we can verify the
tree-level unitarity of the model. But we want spin-$2$ gravitons, so
we redefine $\Lambda$ as $(-\Lambda )$ and, so, for the spin-$2$
sector, we have 
\begin{equation}
k^2=\Lambda \ .
\end{equation}
For the spin-$0$ sector, we have, respectively:
\begin{equation}
k^2=-\ \frac{1}{9}\ \Lambda \ ,
\end{equation}
\[
k^2=\Lambda \ .
\]

Now we have one non tachyonic pole for the spin-$2$ sector, one non
tachyonic and one  tachyonic pole for the spin-$0$ sector.

And we find
\begin{equation}
Im\ Res{\cal A}|_{k^2=\Lambda}>0 \,
\end{equation}

Thus the massive excitation is, in fact, a dynamical degree of
freedom. The theory does not have negative norm states. And this is the
unitarity requisite to have the required assymptotic behavior. From (41),
it is simple to 
verify that the  propagator is proportional to 
$\displaystyle{\frac{1}{k^4}}$, and therefore the $4D$ model is not renormalizable.

So we conclude that the model discussed is causal, has tree-level unitarity
and is not renormalizable by power counting. The proposed model 
has to be understood as an ``effective gravitation theory'' with a
physical massive degree of freedom $(k^2=\Lambda )$.

\subsection*{Acknowledgements:}

\paragraph*{}

The authors want to thank G.O. Pires for his comments and technical
discussions and S.A. Diniz for his work on typing. We also thank
Prof. Berth Schr\"{o}er for many 
discussions during his visit to the Department of Physics of UFES.
We also would like to thank the Department of Physics, University of
Alberta for their hospitality. 
Finally we thank Conselho Nacional de Pesquisa -- CNPq -- for the
financial support.


\begin{thebibliography}{20}
\bibitem{um} K.R. Brownstein, Phys. Rev. A 35,  4854 (1987);
\bibitem{dois} N. Salingaros, J. Phys. A 19, L 101 (1986);
\bibitem{tres} G.F. Freire, Am. J. Phys. 34, 567 (1966);
\bibitem{quatro} C. Chu and T. Ohkawa, Phys. Rev. Lett. 48, 837 (1982);
\bibitem{cinco} K.K. Lee, Phys. Rev. Lett. 50, 138 (1983);
\bibitem{seis} N. Salingaros, Am. J. Phys. 53, 361 (1985);
\bibitem{sete} C. Chu, Phys. Rev. Lett. 50, 139 (1983);
\bibitem{oito} M. Maheswaran, J. Phys. A 19, L761 (1986); 
\bibitem{nove} C. Pinheiro and G.O. Pires, Phys. Letters B. 301 
339 (1993); 
\bibitem{dez} George Arfken, Mathematical Methods for Physicists,
second edition (Academic Press). 
\bibitem{onze} Carlos Pinheiro, G.O. Pires and Nazira Tomimmura, 
General Relativity and Gravitation, vol. 29, N$^{\underline{0}}$ 4, 
409 (1997).  
\bibitem{doze} Carlos Pinheiro, G.O. Pires and C. Sasaki and Il
Nuovo Cimento B, vol. 111B, N$^{\underline{0}}$ 8, 1023-1028 (1996).
\bibitem{treze} Carlos Pinheiro, G.O. Pires and F.A.B.R. de Carvalho
-- Brazilian J. of Phys., vol. 27, 14 (1997).
\end{thebibliography}
\end{document}